\documentclass[
 reprint,nofootinbib,
 amsmath,amssymb,
 aps,
]{revtex4-2}

\usepackage{graphicx}
\usepackage{dcolumn}
\usepackage{bm}
\usepackage{xcolor}
\usepackage[colorinlistoftodos]{todonotes}

\usepackage{soul}
\usepackage{setspace}

\begin{document}

\preprint{APS/123-QED}

\title{Time-Embedded Convolutional Neural Networks for Modeling Plasma Heat Transport}

\author{Mufei Luo${}^{1}$, Charles Heaton${}^{1}$, Yizhen Wang${}^{1}$, Daniel Plummer${}^{1}$, Mila Fitzgerald${}^{1}$, Francesco Miniati${}^{2}$, Sam M. Vinko${}^{1}$ and Gianluca Gregori${}^{1}$}
   
\affiliation{%
 ${}^1$ Department of Physics, University of Oxford, Parks Road, Oxford OX1 3PU, UK
}%
\affiliation{%
 ${}^2$  Mach42, Robert Robinson Avenue, Oxford Science Park, Oxford, OX4 4GP, UK
}%

\begin{abstract}

We introduce a time-embedded convolutional neural network (TCNN) for modeling spatiotemporal heat transport in plasmas, particularly under strongly nonlocal conditions. In our earlier work, the \textit{LMV-Informed Neural Network} (\textit{LINN}) (Luo~{\it et al}., arXiv:2506.16619) combined prior knowledge from the LMV model with kinetic Particle-in-Cell (PIC) data to improve kernel-based heat-flux predictions. While effective under moderately nonlocal conditions, \textit{LINN} produced physically inconsistent kernels in strongly time-dependent regimes due to its reliance on the quasi-stationary LMV formulation. To overcome this limitation, TCNN is designed to capture the coupled evolution of both the normalized heat flux and the characteristic nonlocality parameter using a unified neural architecture informed by underlying physical principles. Trained on fully kinetic PIC simulations, TCNN accurately reproduces nonlocal dynamics across a broad range of collisionalities. Our results demonstrate that the combination of time modulation, coupled prediction, and convolutional depth significantly enhances predictive performance, offering a data-driven yet physically consistent framework for multiscale plasma transport problems.

\end{abstract}

\maketitle
\section{Introduction}

Accurate modeling of heat transport~\cite{Gianluca1,Arran} in plasmas is fundamental to a variety of high-energy-density applications, including inertial confinement fusion~\cite{ICFoverreviw} and astrophysical phenomena~\cite{Yerger,Komarov}. In these systems, the degree of nonlocality is often characterized by the ratio between the electron mean free path, $\lambda_{\rm free} = 4 \pi \varepsilon_0^2 m_e^2 v_{\rm th}^4 / (Z n_e e^4 \ln\Lambda)$~\cite{huba2023plasma}, and the electron temperature gradient length, $L_T = T_e / \nabla T_e$. Here, $T_e$ is the electron temperature, $\varepsilon_0$ is the vacuum permittivity, $m_e$ is the electron mass, $v_{th}=\sqrt{T_e/m_e}$ is the electron thermal velocity, $Z$ is the ion charge state, $n_e$ is the electron number density, $e$ is the elementary charge, and $\ln\Lambda$ is the Coulomb logarithm. 

When $\lambda_{\rm free}/L_T  \ll 0.002$, electron heat conduction can be reliably described by classical, local models such as the Spitzer--Härm (SH) theory~\cite{SH}. However, as this ratio increases, nonlocal effects emerge, leading to the breakdown of traditional hydrodynamic closures~\cite{nonlocalclosure}. In response, several semi-empirical models have been proposed, including the Luciani--Mora--Virmont (LMV)~\cite{LMV} and Schurtz--Nicolaï--Busquet (SNB)~\cite{SNB} formulations, which aim to extend local theory by introducing an empirical nonlocal kernel~\cite{Epperlein,Batishchev2002}. In recent years, diffusion-based formulations have been reconsidered~\cite{Michel2023}. However, such approaches typically rely on the assumption of temporal stationarity~\cite{Chrisment}. Alternatively, empirical flux-limited diffusion schemes are often used in large-scale hydrodynamic modeling~\cite{Yuan}.

Lamy {\it et al.}~\cite{Lamy2022} constructed databases from CHIC hydrodynamic simulations using the classical SNB~\cite{SNB} kernel, which restricted their neural network to reproducing an existing analytic model rather than learning directly from kinetic data. In contrast, our earlier work~\cite{LINN} introduced the \textit{LMV-Informed Neural Network} (\textit{LINN}), a data-driven framework that integrates prior knowledge from the LMV model to learn nonlocal transport kernels. Trained on fully kinetic Particle-in-Cell (PIC) simulation data, \textit{LINN} successfully captures both the nonlocal kernel structure and the associated heat flux under moderately nonlocal conditions, therefore, improves the kernel-based heat flux model. However, as the degree of nonlocality increases, the initial temperature profile undergoes significant temporal evolution. Although \textit{LINN} maintains accurate predictions of the heat flux in such regimes, it yields physically inconsistent kernels. This limitation arises from the model’s adherence to the LMV formulation, which assumes quasi-stationary temperature profiles and therefore does not generalize to strongly time-dependent scenarios. 

To address the challenges of modeling dynamic, strongly nonlocal heat transport, this study proposes a \textit{Time-embedded Convolutional Neural Network} (TCNN) architecture. In contrast to \textit{LINN}, which focuses on predicting the kernel and its associated heat flux, TCNN is designed to capture the spatiotemporal evolution of both the heat flux and the characteristic ratio $\lambda_{\rm free}/L_T$ using a unified set of learnable parameters. This formulation reflects the underlying coupling between the heat flux and characteristic ratio, providing a more coherent and physically consistent modeling framework. At the same time, TCNN retains the foundational principle of nonlocal transport inspired by the LMV model, namely that the heat flux at a given location is influenced by contributions from all spatial points within the system.

The remainder of this paper is organized as follows. Section~\ref{generation-data} details the generation of kinetic data using PIC simulations, with an emphasis on distinguishing the characteristics of the data under local and nonlocal transport regimes. The architectural design of the TCNN is presented in Sec.~\ref{TCNN-Architecture}, followed by an in-depth evaluation of its predictive performance and comparisons with several architectural variants in Sec.~\ref{TCNN-performance}. Finally, Sec.~\ref{conclusion} offers concluding remarks and discusses the broader implications of this work.

\section{Kinetic Data Generation \label{generation-data}}

To investigate heat transport across different regimes, we follow the same approach used previously~\cite{LINN} and employ one-dimensional (1D) PIC simulations using the OSIRIS code~\cite{osiris}, equipped with physically validated binary collision models~\cite{piccollision1, piccollision2}. The system is initialized with a Maxwellian velocity distribution, and the spatial temperature profile is defined as follows
\begin{equation}
T_e(x) = \frac{2T_{e0}(R - 1)}{(R + 1)[1 + \exp(x/L)]} + \frac{2T_{e0}}{R + 1}.
\label{teprofile}
\end{equation}
We set the central temperature to $T_{e0} = 1$ keV, a value characteristic of conduction-zone plasmas in laser-fusion targets, where typical electron temperatures lie between $\sim$50 eV and 2 keV~\cite{suxinghu1}. $R = T_h / T_c$ represents the hot-to-cold temperature ratio, and $L$ defines the transition width between hot and cold regions. This profile establishes a temperature gradient in the box center, with a high-temperature region on the left and a low-temperature region on the right. The corresponding electron density is set to follow $n_e(x) \propto 1/T_e(x)$, approximating an isobaric condition characteristic of laser-heated conduction zones~\cite{suxinghu1}. To vary the degrees of nonlocality, we adjust the background electron density $n_{\rm e0}$, effectively modifying the mean free path $\lambda_{\rm free}$ relative to the temperature gradient length $L_T = 2L(R + 1)/(R - 1)$. This approach enables a controlled transition from classical transport behavior to regimes dominated by strong nonlocal effects. 

We consider a fully ionized plasma composed of high-$Z$ ions, specifically with $Z = 16$ to enhance collisionality. The Coulomb logarithm, $\ln\Lambda$, is computed at the center of the simulation domain and held constant within each run. We employ normalized units based on the plasma frequency $\omega_{\rm pe}=(n_{e0}e^2/\epsilon_0m_e)^{1/2}$ and wavenumber $k_{\rm pe}=\omega_{\rm pe}/c$, with $c$ the light velocity in vacuum. The box length is fixed at $L_{\rm box} \approx 1500$, with spatial resolution $\Delta x = 0.04$ and time step $\Delta t = 0.035$, satisfying Courant–Friedrichs–Lewy condition ($\Delta t / \Delta x \approx 0.9$) and resolving the Debye length ($\lambda_{\rm De,0} = 1.2 \Delta x$). Each cell contains 25,000 electron and 1,600 ion macroparticles. 

\begin{figure}[t]
    \centering
    \includegraphics[width=0.95\linewidth]{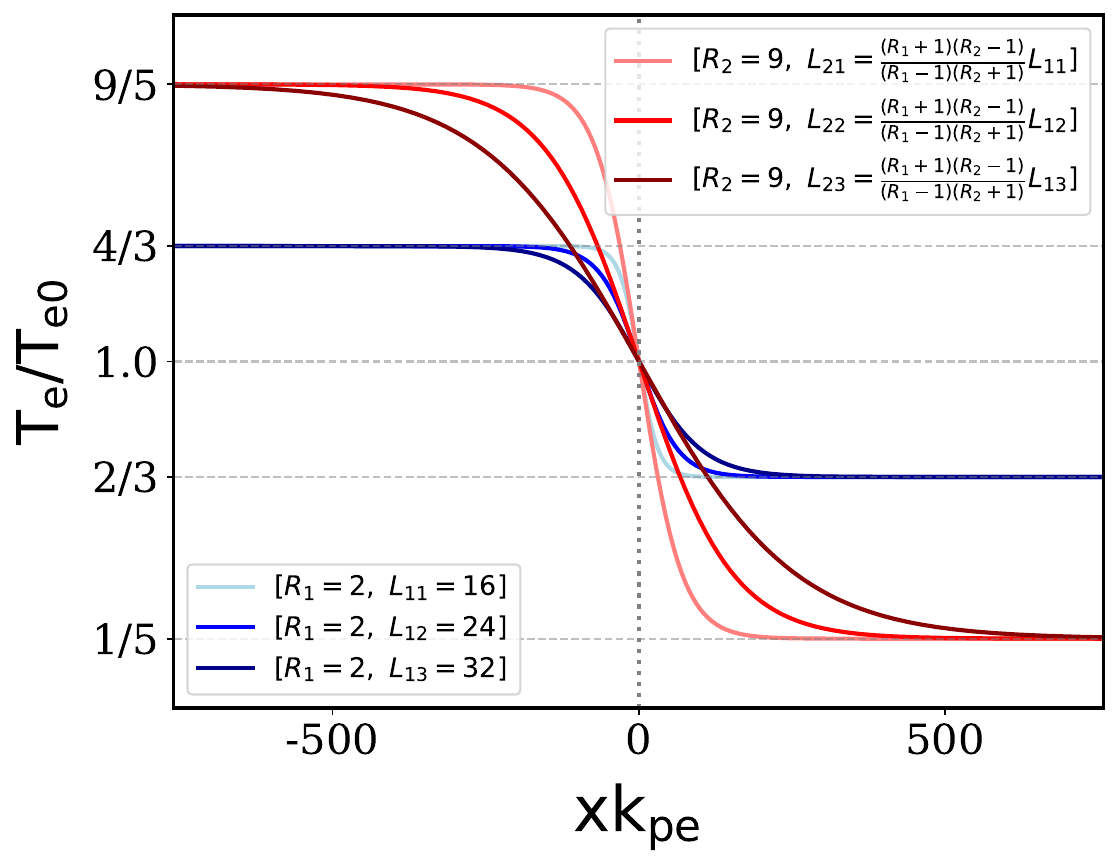}
    \caption{
    Initial electron temperature profiles used to explore the effects of nonlocality and high-energy electrons on heat transport. The light-blue, blue, and dark-blue curves correspond to a fixed temperature ratio of $R_1 = 2$ and increasing transition widths $L = 16$, $24$, and $32$, resulting in temperature gradient lengths of $L_T = 96$, $144$, and $192$ at the center of the domain. To explore the influence of hot electrons while maintaining a constant gradient length, the temperature ratio $R$ is increased from 2 to 9, and the transition width $L$ is tuned according to $L_T = 2L(R + 1)/(R - 1)$, shown by the light-red, red, and dark-red curves. 
    }
    \label{CNN1}
\end{figure}

Figure~\ref{CNN1} illustrates the initial electron temperature profiles used to control the nonlocality. The light-blue, blue, and dark-blue curves correspond to a fixed temperature ratio of $R_1 = 2$, with varying transition widths of $L = 16$, $L = 24$, and $L = 32$, respectively. These values yield corresponding temperature gradient lengths at the center of the domain of $L_T = 96$, $L_T = 144$, and $L_T = 192$.
The electron mean free path $\lambda_{\rm free}$ is changed by adjusting the background electron density $n_{\rm e0}$, thereby modifying the characteristic ratio $\lambda_{\rm free}/L_T$. To examine the influence of hotter electrons on heat transport, we increase the temperature ratio $R = T_h / T_c$ while maintaining a constant temperature gradient length $L_T$ at the center of the domain. This is achieved by appropriately adjusting the transition width $L$ in accordance with the relation $L_T = 2L(R + 1)/(R - 1)$. As shown by the light-red, red, and dark-red curves in Fig.~\ref{CNN1}, these profiles preserve the same gradient length as their corresponding $R=2$ curves at the domain center but exhibit progressively higher temperatures on the hot side. Accordingly, the temperature ratio $R$ is varied from 2 to 9 to systematically assess the impact of increasingly energetic electrons on heat transport. The background electron density is varied across different simulations but is restricted to values not exceeding $10^{25}\,\mathrm{cm^{-3}}$ to ensure that the plasma remains in a non-degenerate regime.

\begin{figure*}
    \centering
    \includegraphics[width=0.95\linewidth]{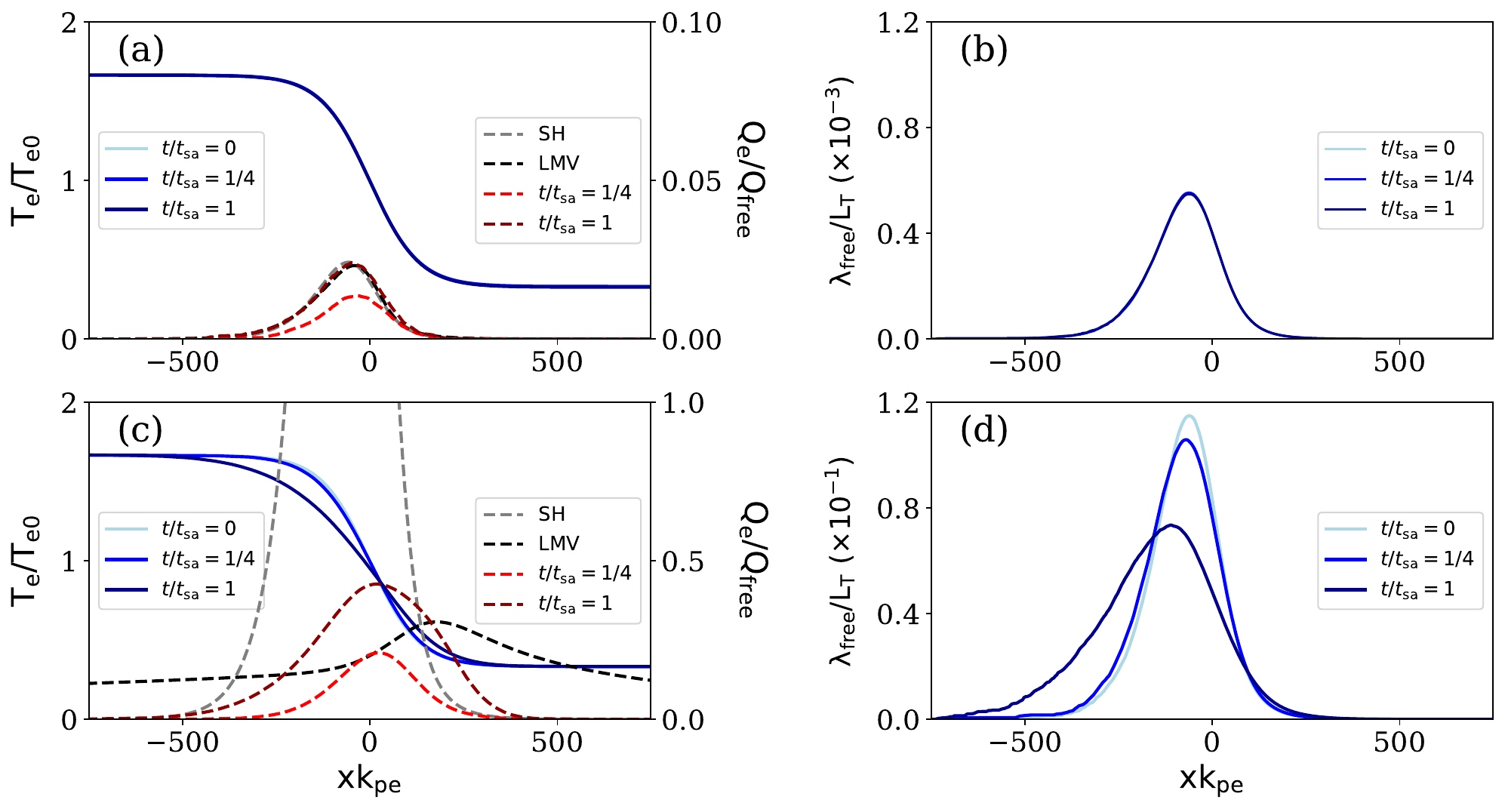}
    \caption{
    Heat transport in the local and strongly nonlocal regimes, both initialized with a temperature ratio of $R = 5$ and gradient length $L_T = 192$. (a) For the local case with central density $n_{\rm e0} = 10^{25}\,\mathrm{cm^{-3}}$ ($\lambda_{\rm free}/L_T \approx 0.0004$), temperature profiles are shown at normalized times $t/t_{\rm sa} = 0$, $1/4$, and $1$ (solid light-blue, blue, and dark-blue curves, left $y$-axis), along with corresponding heat fluxes at $t/t_{\rm sa} = 1/4$ and $1$ (dashed red and dark-red curves, right $y$-axis). Classical Spitzer–Härm (SH) and nonlocal LMV model predictions are shown for comparison (dashed gray and black curves). The saturated heat flux remains consistent with theoretical models and induces negligible temperature evolution. (b) The characteristic ratio $\lambda_{\rm free}/L_T$ remains temporally invariant, as the normalized heat flux, $Q_e/Q_{\rm free}$, is sufficiently small to induce negligible thermal evolution. (c) In the strongly nonlocal case ($n_{\rm e0} = 10^{19}\,\mathrm{cm^{-3}}$, $\lambda_{\rm free}/L_T \approx 0.084$), PIC simulations reveal significant suppression and preheating in the heat flux relative to SH predictions. LMV fails to capture the observed transport behavior. Strong heat flux leads to temporal evolution of the temperature profile and substantial variation in $\lambda_{\rm free}/L_T$ shown in (d).
}
    \label{CNN2}
\end{figure*}

Figure~\ref{CNN2} presents two representative cases illustrating heat transport behavior in the local and nonlocal regimes. Both scenarios are initialized with a temperature ratio of $R = 5$ and a temperature gradient length of $L_T = 192$. In the case shown in Fig.~\ref{CNN2}(a) and (b), the central electron density is set to $n_{\rm e0} = 10^{25}\,\mathrm{cm^{-3}}$, corresponding to a characteristic ratio of $\lambda_{\rm free}/L_T \approx 0.0004$. This places the system well within the collisional (local) transport regime. Figure~\ref{CNN2}(a) displays the evolution of the electron temperature (left $y$-axis) at three distinct normalized times, $t/t_{\rm sa} = 0$, $1/4$, and $1$, represented by the solid light-blue, blue, and dark-blue curves, respectively. $t_{\rm sa}$ denotes the saturation time specific to each case.
The corresponding heat fluxes (right $y$-axis) at $t/t_{\rm sa} = 1/4$ and $1$ are indicated by the dashed red and dark-red curves. For comparison, theoretical predictions based on the SH and LMV models are overlaid using dashed gray and black curves, respectively. 

The classical local heat transport prediction of SH flux~\cite{SH} is given by $Q_{\rm SH}/Q_{\rm free} = a\, \lambda_{\rm SH} / L_T$, where $a = 128(Z + 0.24) / [3\pi(Z + 4.2)]$ and $\lambda_{\rm SH} = 3 \lambda_{\rm free} / \sqrt{\pi/2}$ is the classical Spitzer–Härm mean free path. And the nonlocal LMV model is expressed as~\cite{LMV}
\begin{equation}
    Q_e(x) = \int Q_{\rm SH}(x^{\prime})\, \mathcal{W}(x, x^{\prime})\, dx^{\prime},
    \label{nonlocaltheory}
\end{equation}
the kernel function $\mathcal{W}(x, x')$ quantifies the contribution of the local SH flux at position $x'$ to the net flux at position $x$, thereby encapsulating nonlocal transport effects. It is defined as:

\begin{equation}
    \mathcal{W}(x, x') = \frac{1}{2\lambda(x')} \exp\left[-\frac{X(x,x')}{\lambda(x')}\right],
    \label{theriticalkernel}
\end{equation}
where $X$ represents a density-weighted spatial separation between $x$ and $x'$, computed as:

\begin{equation}
    X = \frac{1}{n_e(x')} \left| \int_x^{x'} n_e(x'')\, dx'' \right|,
    \label{distance}
\end{equation}
and
\begin{equation}
    \lambda(x') \simeq 30\, \sqrt{Z + 1}\, \lambda_{\rm free}(x').
    \label{mfp}
\end{equation}

As shown in Fig.~\ref{CNN2}(a), the saturated heat flux exhibits strong agreement with both the SH and LMV model predictions. The temperature profile remains effectively stationary throughout the simulation, as the normalized heat flux, $Q_e/Q_{\rm free}$, is sufficiently small to induce negligible thermal evolution. Consequently, the characteristic ratio $\lambda_{\rm free}/L_T$ remains stationary throughout the space, as shown in Fig.~\ref{CNN2}(b). 

\begin{figure*}
    \centering
    \includegraphics[width=0.95\linewidth]{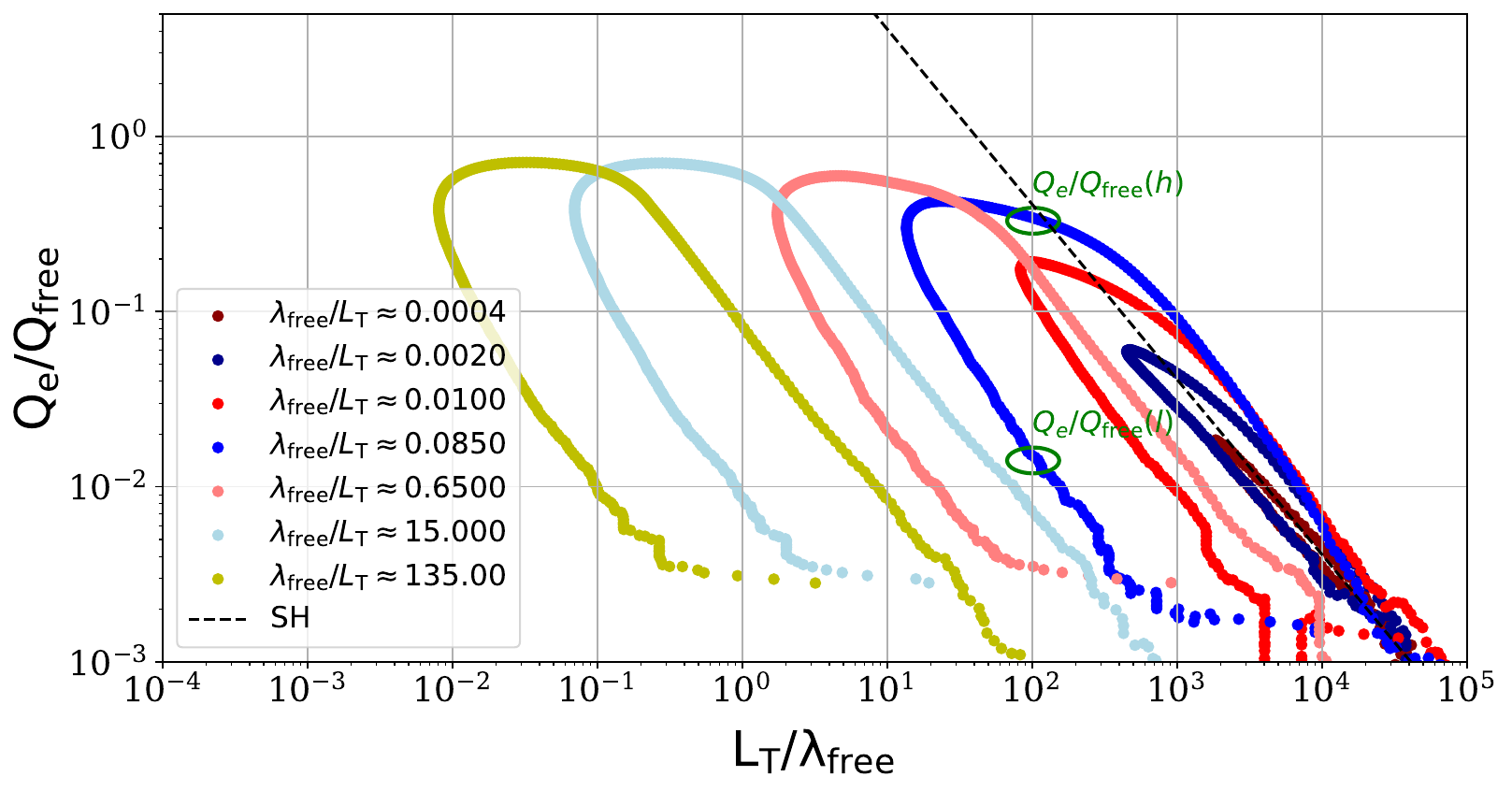}
    \caption{
    Normalized heat flux $Q_e/Q_{\rm free}$ as a function of the inverse characteristic ratio  $L_T/\lambda_{\rm free}$. All cases are initialized with a fixed temperature ratio $R = 5$ and a central temperature gradient length $L_T = 192$, while varying the background electron density. Data points correspond to central densities of $n_{\rm e0} = 10^{25}$ (dark red), $10^{23}$ (dark blue), $10^{21}$ (red), $10^{19}$ (blue), $10^{17}$ (light red), $10^{14}$ (light blue), and $10^{12}\,\mathrm{cm^{-3}}$ (yellow), yielding characteristic ratios of $\lambda_{\rm free}/L_T \approx 0.0004$, 0.002, 0.01, 0.085, 0.65, 15, and 135, respectively. The dashed black line shows the classical SH prediction.
}
    \label{CNN3}
\end{figure*}

In contrast, Fig.~\ref{CNN2}(c) presents a scenario with a much lower central electron density, $n_{\rm e0} = 10^{19}\,\mathrm{cm^{-3}}$, resulting in a characteristic ratio of $\lambda_{\rm free}/L_T \approx 0.084$. This firmly places the system within the strongly nonlocal regime. Under these conditions, the heat flux obtained from the PIC simulation shows a marked suppression and evidence of preheating when compared to the classical SH prediction. Furthermore, the LMV model fails to accurately capture the heat flux dynamics in this regime. Owing to the relatively large normalized heat flux, $Q_e/Q_{\rm free}$, the temperature profile evolves significantly over time. This dynamic behavior also induces substantial variation in the characteristic ratio $\lambda_{\rm free}/L_T$, as illustrated in Fig.~\ref{CNN2}(d). Thus, the onset of strong nonlocality introduces pronounced temporal dynamics into the system, conditions under which our previous model, \textit{LINN}, demonstrates limited effectiveness. It is worth noting that the time-independent heat flux formulation proposed by Epperlein and Short~\cite{Epperlein} exhibits a higher degree of localization compared to the LMV model, particularly within the strongly nonlocal regime; however, it nevertheless underestimates the heat flux relative to kinetic simulations.

To facilitate direct comparison with previous studies on heat transport~\cite{bell,piccollision1,collisonexample3}, Fig.~\ref{CNN3} presents the normalized heat flux, $Q_e/Q_{\rm free}$, plotted against the inverse characteristic ratio, $L_T/\lambda_{\rm free}$. All cases shown share a fixed temperature ratio of $R = 5$ and a central temperature gradient length of $L_T = 192$, while differing in background electron density. The plotted data points correspond to central densities of $n_{\rm e0} = 10^{25}$ (dark-red), $10^{23}$ (dark-blue), $10^{21}$ (red), $10^{19}$ (blue), $10^{17}$ (light-red), $10^{14}$ (light-blue), and $10^{12}\,\mathrm{cm^{-3}}$ (yellow). These density options correspond to the characteristic ratio $\lambda_{\rm free}/L_T\approx$ 0.0004 (dark-red), 0.002 (dark-blue), 0.01 (red), 0.085 (blue), 0.65 (light-red), 15 (light-blue), and 135 (yellow). The dashed black line plots the SH flux.

Within each case examined in Fig.~\ref{CNN3}, for a given value of $L_T/\lambda_{\rm free}$, two distinct values of the normalized heat flux $Q_e/Q_{\rm free}$ are identified: a higher value, $Q_e/Q_{\rm free}(h)$, representing the heat flux in the downstream, and a lower value, $Q_e/Q_{\rm free}(l)$, corresponding to the upstream one. As a representative example, two green ellipses are marked to indicate $Q_e/Q_{\rm free}(h)$ and $Q_e/Q_{\rm free}(l)$ for $L_T/\lambda_{\rm free} = 100$ within the blue scatter data. In the case of $\lambda_{\rm free}/L_T \approx 0.0004$, the system across the space resides in the local transport regime. Both $Q_e/Q_{\rm free}(h)$ and $Q_e/Q_{\rm free}(l)$ closely align with the classical SH theory. However, for $\lambda_{\rm free}/L_T \approx 0.002$, the local condition begins to break down on the hot side of the temperature gradient, where electrons become sufficiently energetic to exhibit nonlocal behavior. These nonlocal electrons propagate toward the colder region, resulting in enhanced heat flux in the downstream region, $Q_e/Q_{\rm free}(h)$ exceeds SH predictions, while the upstream flux $Q_e/Q_{\rm free}(l)$ is diminished. This asymmetry intensifies with increasing nonlocality. As $\lambda_{\rm free}/L_T$ grows, the peak heat flux begins to fall below SH levels, indicating flux suppression due to nonlocal transport. 

Furthermore, once $\lambda_{\rm free}/L_T$ exceeds approximately 0.1, the system transitions into a collisionless regime within the duration of the simulation ($T\omega_{\rm pe} = 2500$), and no saturation state is reached. Consequently, the maximum achievable heat flux plateaus and shows no further enhancement despite increasing nonlocality. For this reason, the subsequent studies are confined to cases where the characteristic ratio satisfies $\lambda_{\rm free}/L_T \leq 0.1$, ensuring that the system remains in a regime where meaningful saturation of the heat flux can be observed. Correspondingly, the considered range of central electron densities spans from $10^{19}$ to $10^{25}\,\mathrm{cm^{-3}}$.

\begin{figure*}
    \centering
    \includegraphics[width=0.85\linewidth]{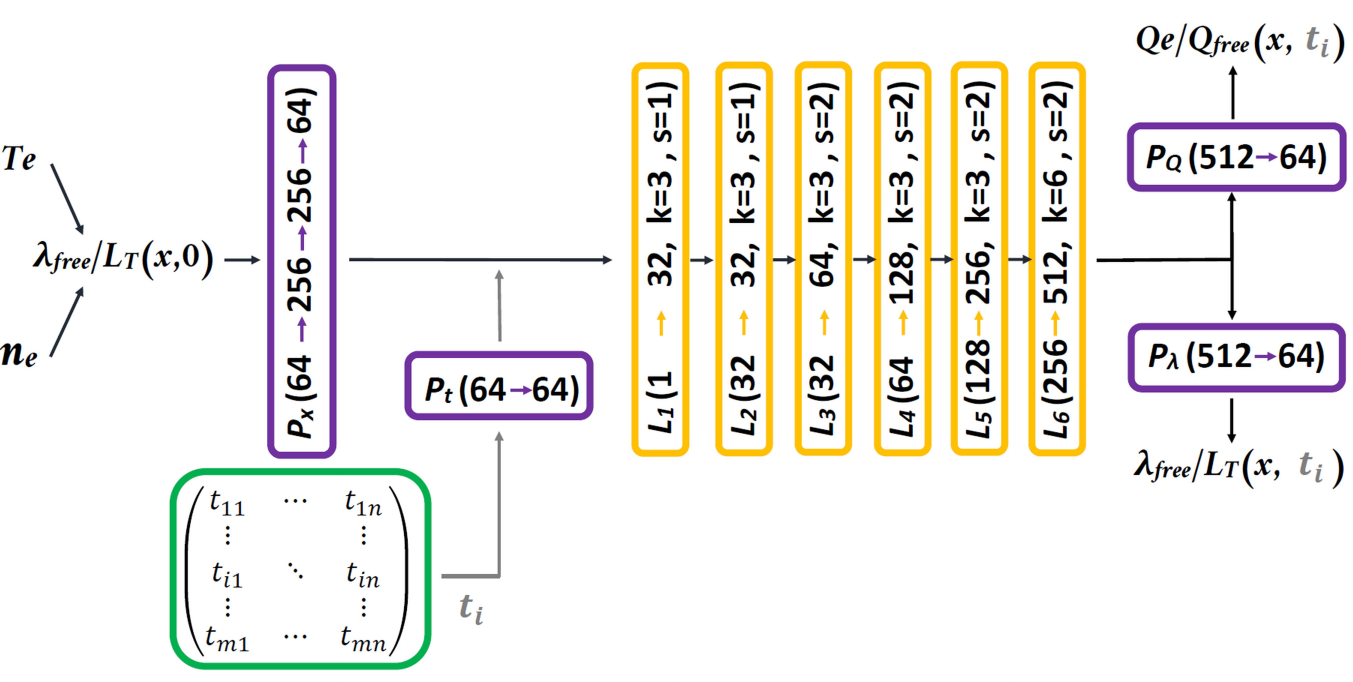}
        \caption{
    \textbf{Architecture of the Time-Embedded Convolutional Neural Network (TCNN).} The model predicts spatiotemporal quantities, characteristic ratio $\lambda_{\rm free}/L_T(x,t)$ and normalized heat flux $Q_e / Q_{\text{free}}(x, t)$, conditioned on the spatial profile $\lambda_{\text{free}} / L_T(x, 0)$ derived from electron temperature $T_e$ and density $n_e$. The spatial input is projected via a fully connected encoder $P_x$, while the time index is embedded and projected through $P_t$. These are combined and passed through a sequence of 6 1D convolutional layers $L_1$--$L_6$ with progressively increasing channels. Two projection heads $P_\lambda$ and $P_Q$ decode the latent representation into physically meaningful outputs. The time matrix, highlighted within the green box, has dimensions \( (n, m) \), where \( n=64 \) and \( m=44 \) denote the number of spatial and temporal steps, respectively.
    }
    \label{TCNNmodel}
\end{figure*}

\section{Time-Embedded Convolutional Neural Networks Architecture  \label{TCNN-Architecture}}

In our previous work~\cite{LINN}, the \textit{LINN} model was designed to incorporate as much theoretical structure from the LMV model as possible. Specifically, the elements of the nonlocal kernel $\mathcal{W}$ (Eq.~(\ref{theriticalkernel})) were generated by mapping structured physical inputs -- namely $X$ (Eq.~(\ref{distance})) and $\lambda$ (Eq.~(\ref{mfp})) -- through the neural network architecture of \textit{LINN}. The resulting kernel was then used to compute the nonlocal heat flux by convolving the classical SH flux $Q_{\rm SH}$ over space, as prescribed by Eq.~(\ref{nonlocaltheory}). The SH flux itself was evaluated using the initial value of the characteristic ratio $\lambda_{\rm free}/L_T$, maintaining consistency with the LMV-based nonlocal transport formulation. 

However, as demonstrated in Fig.~\ref{CNN2}(c) and (d), under strongly nonlocal conditions, the normalized heat flux $Q_e/Q_{\rm free}$ becomes sufficiently large enough to drive a substantial evolution of the temperature profile. Consequently, the characteristic ratio $\lambda_{\rm free}/L_T$ also undergoes temporal variation. In such dynamic scenarios, relying on the initial value of $\lambda_{\rm free}/L_T$ to compute the nonlocal heat flux yields physically inconsistent kernels, though it can still maintain accurate predictions of the heat flux. From a physical perspective, the intrinsic coupling between the heat flux and the parameter $\lambda_{\rm free}/L_T$ suggests that both quantities should be predicted simultaneously using a unified set of parameters within a neural network framework. To achieve this objective, we introduce a Time-Embedded Convolutional Neural Network (TCNN) framework, presented in the following.

In contrast to \textit{LINN}, which strictly adheres to the formalism of Eq.~(\ref{nonlocaltheory}), the TCNN architecture relaxes this constraint to enable the discovery of alternative, data-driven representations of nonlocal heat transport. Nonetheless, it preserves the core physical principle that the heat flux at any given location is influenced by contributions from the entire spatial domain. Specifically, TCNN is designed such that the predicted heat flux at a point incorporates information about the spatial distribution of the parameter $\lambda_{\rm free}/L_T$. We rewrite Eq.~(\ref{nonlocaltheory}) as 
\begin{equation}
    Q_e(x,t)/Q_{\rm free,0} = \int \widetilde{a}(x,x^{\prime},t) \ \lambda_{\rm free} (x^{\prime})/ L_T(x^{\prime})\, \, dx^{\prime}.
    \label{TCNNidea}
\end{equation}
Here, $Q_{\rm free,0}$ denotes the free-streaming heat flux evaluated at the center of the domain, while temporal information is implicitly encoded within the learned $\widetilde{a}$.

Inspired by Eq.~(\ref{TCNNidea}), several NN modeling strategies can be proposed in order of complexity:

\begin{enumerate}
    \item Operating directly in real space by applying a single convolutional layer to convolve parameter $\lambda _{\rm free}/L_T$ across the domain;

    \item Lifting the input vector ($\lambda _{\rm free}/L_T$) into a latent space, and applying a single convolutional layer to convolve the lifted vector across the latent domain;

    \item Operating directly in real space, but applying multiple convolutional layers to convolve parameter $\lambda _{\rm free}/L_T$ layer by layer across the domain;

    \item Lifting the input vector ($\lambda _{\rm free}/L_T$) into a latent space, before applying multiple convolutional layers to convolve the lifted vector across the latent domain.
\end{enumerate}
Furthermore, in any of the above strategies, the temporal information must be introduced to capture the dynamics of heat transport.. The first strategy directly adheres to the formulation of Eq.~(\ref{TCNNidea}). However, in the context of nonlocal transport, the inherent nonlocal characteristics of the system could be more effectively captured within one latent space~\cite{no}, which the model needs to learn. Furthermore, the use of multiple convolutional layers can enhance the model’s ability to capture nonlocal and nonlinear features of the system. This aspect will be further elucidated in the following.

Figure~\ref{TCNNmodel} shows the resulting TCNN architecture that captures both spatial correlations and temporal dependencies through explicit time embedding. 
The model operates on the initial profile of the characteristic ratio $\lambda_{\rm free}/L_T$, which is discretized into a 64-point grid.
It is conditioned on a discrete time index $t_i/t_{\rm sa} \in \{1/44, 2/44, \dots, 1\}$, corresponding to one of 44 time steps (see the time matrix within the green box, the elements of this time matrix are trainable).

A learnable embedding layer encodes the discrete time step $t_i$ by delivering the $i$-th row of the time matrix to another trainable vector in $\mathbb{R}^{64}$ (see projection labeled '$P_t$'). This is projected through a fully connected (FC) network into a 64-dimensional latent space to match the spatial feature embedding, allowing time-aware conditioning of the model.

The spatial input vector $\lambda_{\rm free}/L_T \in \mathbb{R}^{64}$ is first passed through an FC network (see projection labeled '$P_x$'):
\[
\lambda_{\rm free}/L_T \rightarrow \text{FC}(64 \rightarrow 256) \rightarrow \text{GELU} \rightarrow \text{FC}(256 \rightarrow 64)
\]
to embed it into a latent feature representation. The time-conditioned embedding is added to this output to yield a time-aware spatial representation.

Subsequently, the resulting feature is processed through a series of 6 1D convolutional layers to extract hierarchical spatial features, shown by the yellow boxes in Fig.~\ref{TCNNmodel}. And GELU activations are applied after each layer.

The final latent representation is passed through two separate linear heads to predict two target quantities (see projections labeled '$P_\lambda$' and '$P_Q$'):
\begin{itemize}
  \item $\lambda_{\text{free}} / L_T(x, t_i)$: the evolved characteristic ratio at time $t_i$;
  \item $Q_e / Q_{\text{free}}(x, t_i)$: normalized heat flux at time $t_i$.
\end{itemize}

As illustrated in the flowchart of Fig.~\ref{TCNNmodel}, the core mechanism for learning the system’s dynamics involves embedding temporal information by introducing distinct time vectors into the initial conditions. Conditioned on the initial characteristic ratio profile, $\lambda_{\rm free}/L_T$, the introduced time vector serves to temporally modulate the convolutional layer parameters. With the exception of the final two FC layers ($P_Q$ and $P_\lambda$), the model employs a largely shared set of parameters to jointly predict the evolution of both $\lambda_{\rm free}/L_T$ and the normalized heat flux $Q_e/Q_{\rm free}$ — thereby preserving the inherent physical coupling between nonlocality and heat transport. Following the convolutional layers, the spatial domain is progressively reduced such that all spatial points are aggregated into a single point, while the feature dimension expands into a latent representation of 512 channels. This architectural design is consistent with the nonlocal LMV model, wherein the heat flux at any given location receives contributions from the entire spatial domain.

In addition to the model architecture, several strategies are required to address the imbalance in data distribution. As illustrated in Fig.~\ref{CNN2} and Fig.~\ref{CNN3}, the values of $\lambda_{\rm free}/L_T$ span approximately two orders of magnitude across the transition from local to strongly nonlocal regimes, while the normalized heat flux $Q_e/Q_{\rm free}$ varies by about one order of magnitude. Notably, $\lambda_{\rm free}/L_T$ remains relatively stationary in the local regime but exhibits pronounced temporal dynamics under strongly nonlocal conditions. These disparities in scale and behavior necessitate tailored techniques to ensure stable and effective learning across regimes.

To address the challenges posed by uneven data distribution across local and nonlocal transport regimes, we introduce a weighted loss function defined as:
\[
\mathcal{L}_{\lambda, Q} = \left\langle \frac{1}{N_b} \sum_{i=1}^{N_b} \frac{(y_i - \hat{y}_i)^2}{(\max(y_i) + \varepsilon)^{p_1}} \right\rangle_x
\]
where \( y \) and \( \hat{y} \) denote the ground truth and predicted values, respectively. \( \mathcal{L}_\lambda \) corresponds to the loss for the characteristic  ratio \( \lambda_{\rm free}/L_T \), and \( \mathcal{L}_Q \) for the normalized heat flux \( Q_e/Q_{\rm free} \). The exponent \( p_1 \) adjusts the scaling of the loss to reflect the magnitude differences across regimes (e.g., \( p_1 = 1.0 \) for \( \lambda_{\rm free}/L_T \), and \( p_1 = 0.75 \) for \( Q_e/Q_{\rm free} \)). The constant \( \varepsilon=10^{-8} \) prevents division by zero, and \( N_b = 16 \) is the patch size. Spatial averaging is applied to compute the loss, represented by the operator \(\langle \, \rangle_{x}\).

To further encourage accurate modeling of temporal dynamics, we incorporate a trend-consistency loss that penalizes deviations in the first-order differences between the predicted and ground truth sequences:
\[
\mathcal{L}_{\text{trend}, (\lambda, Q)} = \left\langle \frac{1}{N_b} \sum_{t=t_2}^{t_{\rm sa}} \left( \Delta \hat{y}_t - \Delta y_t \right)^2 \right\rangle_{x, t}
\]
\[
\Delta y_t = y_{t_i} - y_{t_{i-1}}, \quad \Delta \hat{y}_t = \hat{y}_{t_i} - \hat{y}_{t_{i-1}}
\]
This term is averaged over both spatial and temporal dimensions.

Additionally, to balance variations in the rate of change across regimes, we introduce a differential weighting loss based on the magnitude of changes in the ground truth:
\[
\mathcal{L}_{\text{diff}, (\lambda, Q)} = \left\langle \frac{1}{N_b} \sum_{t=t_2}^{t_{\rm sa}} \frac{(\hat{y}_{t} - \hat{y}_{t-1})^2}{(|y_t - y_{t-1}| + \varepsilon)^{p_2}} \right\rangle_{x, t}
\]
with \( p_2 = 0.8 \) for both \( \lambda_{\rm free}/L_T \) and \( Q_e/Q_{\rm free} \), and averaging over space and time.

Combining these components, the total loss function for a given sequence is:
\begin{align}
\mathcal{L} &= \frac{1}{m} \sum_{t=t_1}^{t_{\rm sa}} \left( 0.5\, \mathcal{L}_\lambda + 0.5\, \mathcal{L}_Q \right) \nonumber \\
&\quad + \mathcal{L}_{\text{trend,$\lambda$}}+ \mathcal{L}_{\text{trend,Q}} \nonumber \\
&\quad + 0.05\,(\mathcal{L}_{\text{diff,$\lambda$}}+\mathcal{L}_{\text{diff,Q}}),
\label{loss}
\end{align}
where \( m \) denotes the total number of time steps. The model is trained using the Adam optimizer with an initial learning rate of 0.001. The \texttt{ReduceLROnPlateau} strategy monitors the loss in \texttt{min} mode and reduces the learning rate by a factor of 0.8 if no improvement is observed for 5 consecutive epochs.

\begin{figure}
    \centering
    \includegraphics[width=1\linewidth]{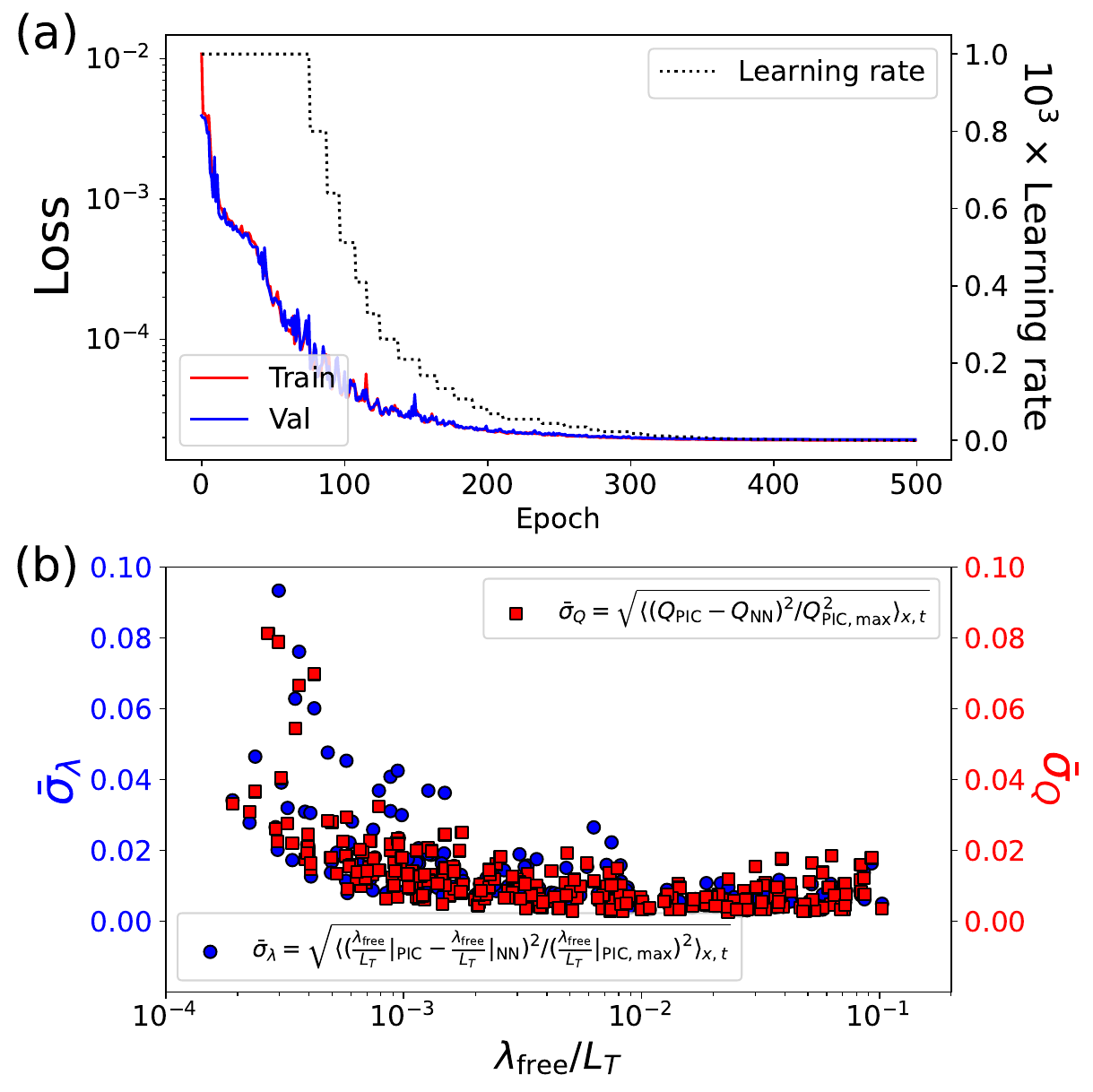}
    \caption{Training and evaluation performance of the TCNN model shown in Fig.~\ref{TCNNmodel}. (a) Training and validation loss curves exhibit smooth convergence, indicating stable optimization without signs of overfitting. (b) Relative prediction errors on the test set, normalized by the maximum values of $Q_e/Q_{\rm free}$ and $\lambda_{\rm free}/L_T$ for each sample, where red markers represent the deviation in normalized heat flux \( Q_e/Q_{\rm free} \) and blue markers indicate the deviation in the characteristic ratio \( \lambda_{\rm free}/L_T \). The horizontal axis denotes the initial value of \( \lambda_{\rm free}/L_T \) in the box center for each test case.}
    \label{trainprocess}
\end{figure}

\begin{figure}
    \centering
    \includegraphics[width=1\linewidth]{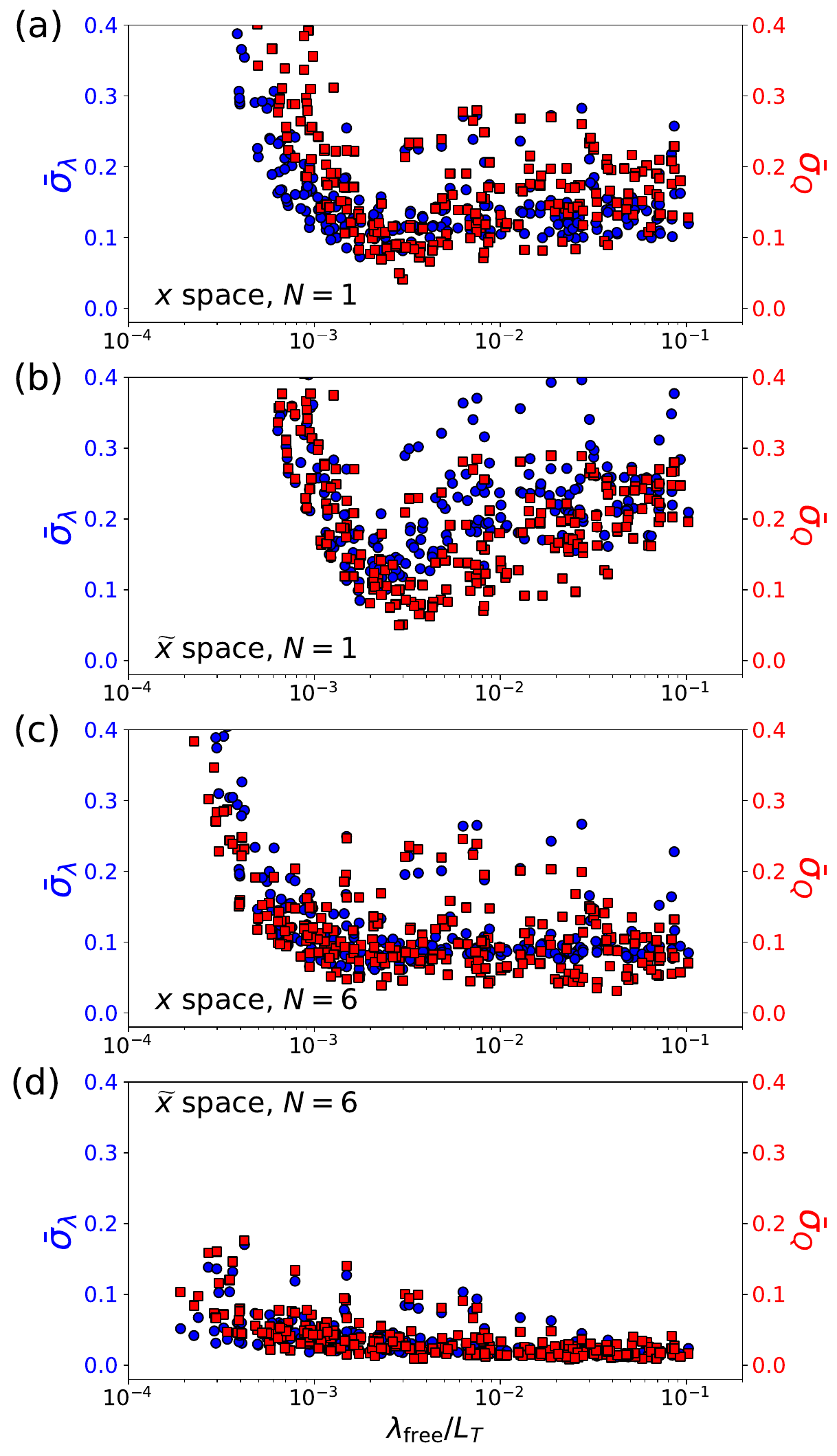}
    \caption{Performance comparison of four architectural variants related to the TCNN framework. (a) A baseline model performing convolution directly in real space without spatial projection, using a single convolutional layer with a kernel size of 64. (b) Similar to (a), but includes a lightweight spatial projection layer \( P_x = \text{FC}(64 \rightarrow 64) \), transforming inputs into a latent space \( \widetilde{x} \). (c) A deeper network with 6 convolutional layers operating in real space. (d) Combining both a lightweight spatial projection layer and 6 convolutional layers.}
    \label{another}
\end{figure}

\section{Performance of TCNN \label{TCNN-performance}}

We utilize the same dataset as in our previous study~\cite{LINN}, partitioned into training, validation, and test subsets in a $70{:}15{:}15$ ratio. To reduce numerical noise and limit the complexity of the neural network, the PIC results comprising 38,400 cells are coarse-grained to 64 sampling points, while still preserving the heat-flux profile. The training process is depicted in Fig.~\ref{trainprocess}(a), which shows the evolution of training and validation losses, calculated via Eq.~(\ref{loss}), for the TCNN model. Both curves exhibit smooth convergence toward a plateau, indicating stable optimization and no evidence of overfitting.

Figure~\ref{trainprocess}(b) assesses the model's predictive performance on the test set without being weighted. Red markers represent the relative deviation between the predicted and ground-truth heat flux values, while blue markers correspond to the deviation in the characteristic ratio, \( \lambda_{\rm free}/L_T \). These deviations are quantified as follows:
\[
\bar{\sigma}_{Q} = \frac{\sqrt{\left\langle \left(Q_{\rm PIC} - Q_{\rm NN}\right)^2 \right\rangle_{x,t}}}{Q_{\rm PIC,\, \mathrm{max}}},
\]
and
\[
\bar{\sigma}_{\lambda} = \frac{\sqrt{\left\langle \left( \left.\frac{\lambda_{\rm free}}{L_T} \right|_{\rm PIC} - \left. \frac{\lambda_{\rm free}}{L_T} \right|_{\rm NN} \right)^2 \right\rangle_{x,t}}}{\left.\frac{\lambda_{\rm free}}{L_T}\right|_{\rm PIC,\, \mathrm{max}}}
\]
Here, \( Q_{\rm PIC,\, \mathrm{max}} \) and \( \left.\lambda_{\rm free}/L_T\right|_{\rm PIC,\, \mathrm{max}} \) denote the maximum values of the respective PIC-based quantities for each sample. 

The horizontal axis in Fig.~\ref{trainprocess}(b) represents the value of \( \lambda_{\rm free}/L_T \) at the center of each test sample. Results indicate that TCNN performs well in capturing the system's spatiotemporal evolution under nonlocal conditions. However, its predictive accuracy tends to degrade in the local regime. This degradation arises because, under highly collisional conditions, the system evolves slowly, and \( \lambda_{\rm free}/L_T \) remain nearly stationary in particular. Consequently, the time-embedded neural architecture receives weak temporal signals, limiting its ability to update and adapt effectively in these near-static regimes. In contrast, our previous model, \textit{LINN}~\cite{LINN}, demonstrated superior predictive capability in the local regime.

\begin{figure}
    \centering
    \includegraphics[width=1\linewidth]{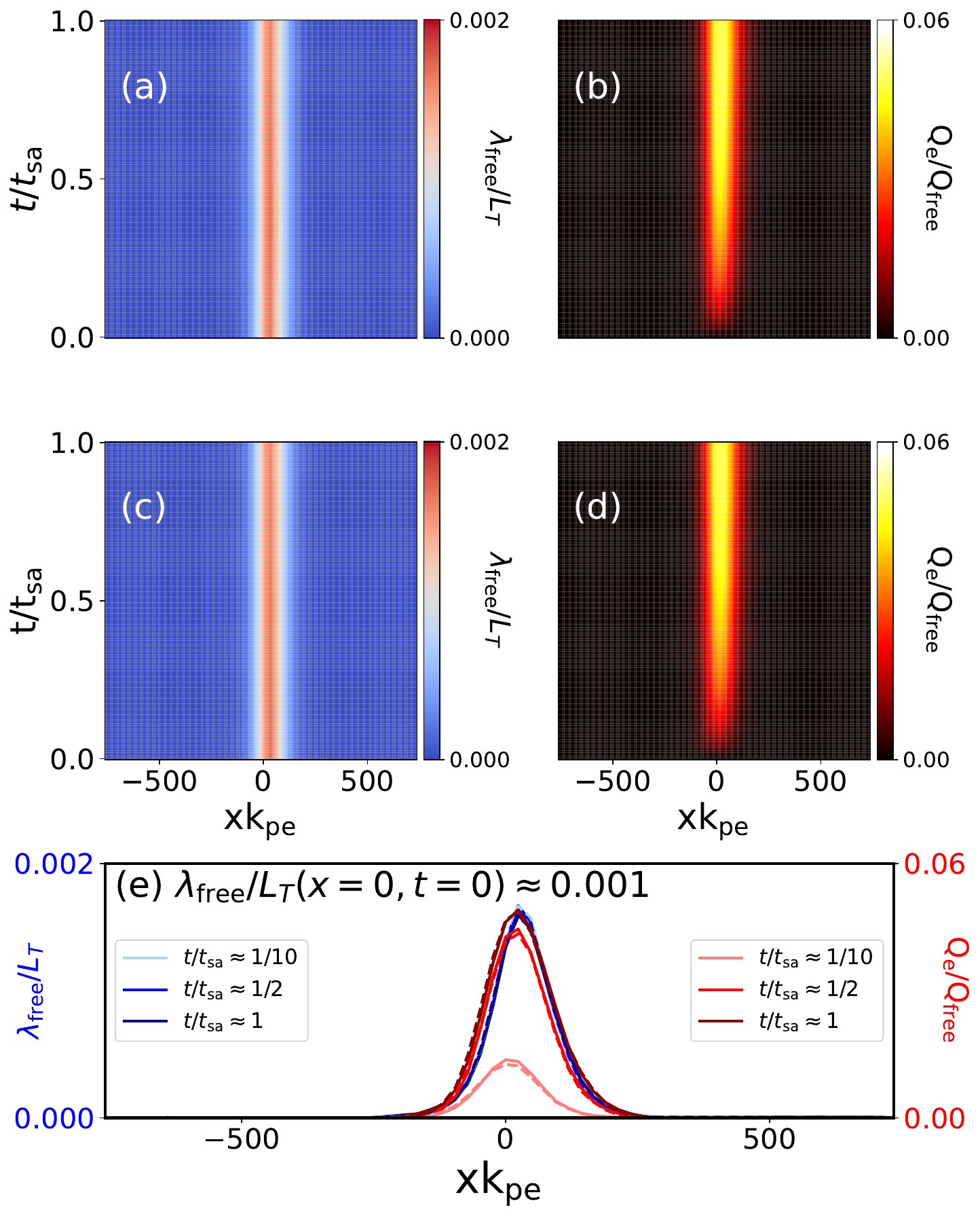}
    \caption{Spatiotemporal prediction results by TCNN in the local regime with an initial characteristic  ratio of \( \lambda_{\rm free}/L_T \approx 0.001 \). (a) and (b) show the model-predicted evolution of \( \lambda_{\rm free}/L_T \) and normalized heat flux \( Q_e/Q_{\rm free} \), respectively. (c) and (d) display the corresponding ground-truth quantities obtained from PIC simulations. (e) presents a direct comparison of spatial profiles at three representative normalized times: \( t/t_{\rm sa} = 1/10 \), \( 1/2 \), and \( 1 \). The solid light-blue, blue, and dark-blue curves represent the TCNN-predicted spatial profiles of \( \lambda_{\rm free}/L_T \) (left \( y \)-axis), while the solid light-red, red, and dark-red curves denote the corresponding predictions for the normalized heat flux \( Q_e/Q_{\rm free} \) (right \( y \)-axis). The dashed curves of matching colors indicate the ground truth results obtained from PIC simulations.}
    \label{modelexample1}
\end{figure}

\begin{figure}
    \centering
    \includegraphics[width=1\linewidth]{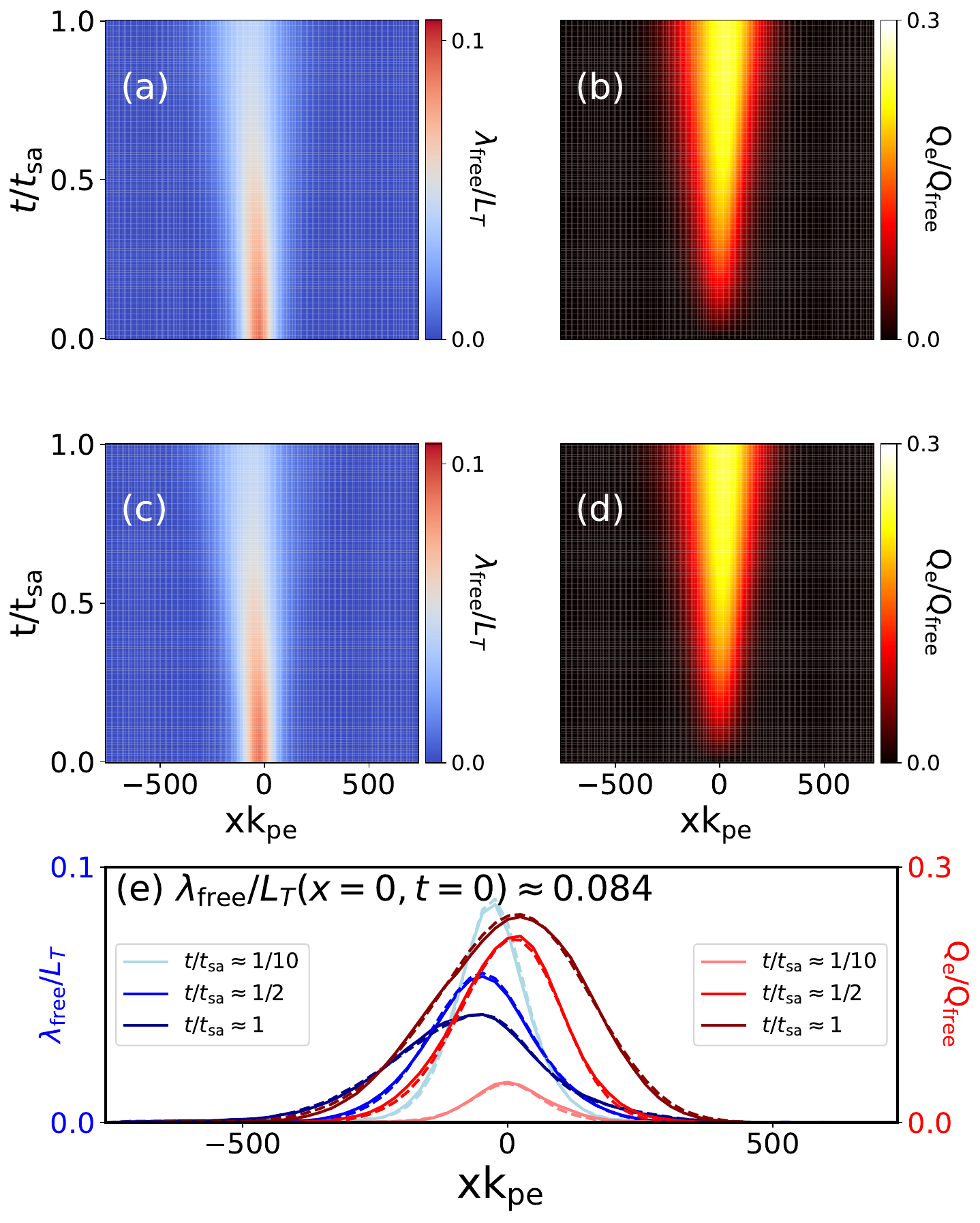}
    \caption{Spatiotemporal prediction results by TCNN in the strongly nonlocal regime with an initial characteristic  ratio of \( \lambda_{\rm free}/L_T \approx 0.084 \). (a) and (b) show the model-predicted evolution of \( \lambda_{\rm free}/L_T \) and normalized heat flux \( Q_e/Q_{\rm free} \), respectively. (c) and (d) display the corresponding ground-truth quantities obtained from PIC simulations. (e) presents a direct comparison of spatial profiles at three representative normalized times: \( t/t_{\rm sa} = 1/10 \), \( 1/2 \), and \( 1 \). The solid light-blue, blue, and dark-blue curves represent the TCNN-predicted spatial profiles of \( \lambda_{\rm free}/L_T \) (left \( y \)-axis), while the solid light-red, red, and dark-red curves denote the corresponding predictions for the normalized heat flux \( Q_e/Q_{\rm free} \) (right \( y \)-axis). The dashed curves of matching colors indicate the ground truth results obtained from PIC simulations.}
    \label{modelexample2}
\end{figure}

The performance of four additional architectural variants, discussed following Eq.~(\ref{TCNNidea}), is summarized in Fig.~\ref{another}. Fig.~\ref{another}(a) illustrates a baseline model in which convolution is performed directly in real space without applying a spatial projection layer ($P_x$). A single convolutional layer with a kernel size of 64 is employed, allowing full spatial convolution. This setup maps the input from 1 channel to 5210 output features, yielding a comparable parameter count (332,800 parameters in the convolutional layer, and 655,488 in the output projections) to the full TCNN architecture (919,648 parameters in the convolutional layer, and 65,664 in the output projections) shown in Fig.~\ref{TCNNmodel}. In Fig.~\ref{another}(b), the architecture builds on (a) by incorporating a simple spatial projection layer $P_x = \text{FC}(64 \rightarrow 64)$, effectively transforming the input into a latent spatial framework $\widetilde{x}$ but maintaining lower expressive capacity than the projection structure in Fig.~\ref{TCNNmodel}. Figure~\ref{another}(c) reverts to real space, but extends the convolutional depth to 6 convolutional layers, identical in depth and structure to those used in the TCNN model depicted in Fig.~\ref{TCNNmodel}. Finally, Fig.~\ref{another}(d) combines both architectural enhancements: a lightweight projection layer $P_x = \text{FC}(64 \rightarrow 64)$ and a 6-layer convolutional stack. 

The architectural variants depicted in Fig.~\ref{another}(a) and (b) exhibit poor performance due to insufficient convolutional depth and limited spatial transformation capabilities, which constrain their ability to model the intrinsic nonlinearity and nonlocality of heat transport. In contrast, the architecture in Fig.~\ref{another}(c) demonstrates notable performance gains through the use of a deeper convolutional network. This capability is further significantly improved in Fig.~\ref{another}(d), where convolution is performed within a latent spatial framework, enabling more expressive feature extraction and better generalization across regimes, although its performance remains slightly inferior to the full TCNN model shown in Fig.~\ref{TCNNmodel}, whose performance is quantified in Fig.~\ref{trainprocess}(b), owing to the latter’s more expressive latent space.

Finally, Fig.~\ref{modelexample1} and Fig.~\ref{modelexample2} showcase two representative examples of the spatiotemporal predictions made by the TCNN model. Figure~\ref{modelexample1} corresponds to a scenario within the local regime, characterized by an initial characteristic ratio of \( \lambda_{\rm free}/L_T \approx 0.001 \), while Fig.~\ref{modelexample2} illustrates a strongly nonlocal case with \( \lambda_{\rm free}/L_T \approx 0.084 \). Panels (a) and (b) in each figure depict the model-predicted evolution of \( \lambda_{\rm free}/L_T \) and the normalized heat flux \( Q_e/Q_{\rm free} \), respectively, across space and time. For comparison, panels (c) and (d) display the corresponding ground truth data generated via PIC simulations. Panel (e) of each figure presents spatial profiles of both quantities at three distinct normalized times, \( t/t_{\rm sa} = 1/10 \), \( 1/2 \), and \( 1 \). The solid light-blue, blue, and dark-blue curves represent the TCNN-predicted spatial profiles of \( \lambda_{\rm free}/L_T \) (left \( y \)-axis), while the solid light-red, red, and dark-red curves denote the corresponding predictions for the normalized heat flux \( Q_e/Q_{\rm free} \) (right \( y \)-axis). The dashed curves of matching colors indicate the ground truth results obtained from PIC simulations. Across both the collisional and strongly nonlocal regimes, the TCNN model accurately reconstructs the spatial and temporal evolution of key transport quantities. 

\section{Conclusion and Discussion \label{conclusion}}

In this study, we presented a novel deep learning framework, the \textit{Time-embedded Convolutional Neural Network} (TCNN), for modeling spatiotemporal heat transport in plasmas, particularly within the strongly nonlocal regime. This architecture is physics-informed, incorporating the underlying principle that heat flux and nonlocality are intrinsically coupled, while allowing data-driven flexibility by relaxing the strict kernel-based formalism of traditional models such as the LMV theory.

A key feature of TCNN is its ability to jointly predict the normalized heat flux \( Q_e/Q_{\rm free} \) and the characteristic  ratio \( \lambda_{\rm free}/L_T \) using a unified set of convolutional parameters. This formulation reflects the underlying physical coupling between heat flux and nonlocality, providing a more coherent and physically consistent modeling framework. By embedding temporal information directly into the model through time-indexed convolutions, TCNN learns to represent dynamic system evolution. This time-embedded modeling allows for capturing the progression of heat transport processes across varying regimes of collisionality, outperforming prior models such as SH, LMV, especially under strongly nonlocal conditions. Although its accuracy degrades slightly in the local, quasi-static case due to diminished temporal variation, this limitation could be addressed by employing a hybrid of the TCNN and \textit{LINN}~\cite{LINN} models. While integration into radiation–hydrodynamics codes would follow the same approach as LMV or SH models, implementation is not our immediate goal. Our current focus is on developing a reliable, physics-informed, and stable neural network to establish the model, while exploring multiscale strategies, such as partitioning hydrodynamic domains into smaller grids treatable with PIC, to overcome scale limitations for future coupling with hydrodynamic codes.

We conducted an extensive evaluation of multiple architectural variants to dissect the impact of convolutional depth and latent spatial projections. Results show that shallow or unprojected models struggle to capture the complexity of nonlocal dynamics, while deeper convolutional stacks operating in latent space substantially improve performance. The full TCNN architecture, with its rich latent representation and depth, achieved sufficient accuracy, highlighting the importance of flexible and extensible design in deep learning models for physical systems.

Beyond the immediate application to plasma heat transport, our methodology offers broader inspirations for applying neural networks to complex physical systems~\cite{stefan1,stefan2,mfluo1,mfluo2}. In particular, the integration of temporal modulation, joint multi-target prediction, and architecture-informed loss functions provides a practical framework for modeling multiscale, nonlinear, and nonlocal processes. We hope this work serves as a reference for researchers seeking to design models that are both physically grounded and data-efficient, and helps bridge the gap between traditional physics-based modeling and modern machine learning~\cite{miniati, UCLA, fno, Carvalho2024,McDevitt,Dopp2023,Ingelsten}.

The authors thank S. Hu, S. Nathaniel, S. Hüller, T. Fülöp, and C. Riconda for their useful discussions. They are also grateful for the computing resources provided by the STFC Scientific Computing Department’s SCARF cluster. This work was supported by EPSRC and First Light Fusion under the AMPLIFI prosperity partnership, Grant No. EP/X025373/1.

\bibliography{reference}

\end{document}